\documentclass[]{revtex4}
\usepackage{graphicx}
\usepackage{fancyhdr}

\begin{document}

\def\ds{\displaystyle}
\def\beq{\begin{equation}}
\def\eeq{\end{equation}}
\def\bea{\begin{eqnarray}}
\def\eea{\end{eqnarray}}
\def\beeq{\begin{eqnarray}}
\def\eeeq{\end{eqnarray}}
\def\ve{\vert}
\def\vel{\left|}
\def\ver{\right|}
\def\nnb{\nonumber}
\def\ga{\left(}
\def\dr{\right)}
\def\aga{\left\{}
\def\adr{\right\}}
\def\lla{\left<}
\def\rra{\right>}
\def\rar{\rightarrow}
\def\nnb{\nonumber}
\def\la{\langle}
\def\ra{\rangle}
\def\ba{\begin{array}}
\def\ea{\end{array}}
\def\tr{\mbox{Tr}}
\def\ssp{{\Sigma^{*+}}}
\def\sso{{\Sigma^{*0}}}
\def\ssm{{\Sigma^{*-}}}
\def\xis0{{\Xi^{*0}}}
\def\xism{{\Xi^{*-}}}
\def\qs{\la \bar s s \ra}
\def\qu{\la \bar u u \ra}
\def\qd{\la \bar d d \ra}
\def\qq{\la \bar q q \ra}
\def\gGgG{\la g^2 G^2 \ra}
\def\q{\gamma_5 \not\!q}
\def\x{\gamma_5 \not\!x}
\def\g5{\gamma_5}
\def\sb{S_Q^{cf}}
\def\sd{S_d^{be}}
\def\su{S_u^{ad}}
\def\rl{\hat{m}_{\ell}}
\def\ss{\hat{s}}
\def\rr{\hat{r}_{K_1}}
\def\sbp{{S}_Q^{'cf}}
\def\sdp{{S}_d^{'be}}
\def\sup{{S}_u^{'ad}}
\def\ssp{{S}_s^{'??}}
\def\sig{\sigma_{\mu \nu} \gamma_5 p^\mu q^\nu}
\def\fo{f_0(\frac{s_0}{M^2})}
\def\ffi{f_1(\frac{s_0}{M^2})}
\def\fii{f_2(\frac{s_0}{M^2})}
\def\O{{\cal O}}
\def\sl{{\Sigma^0 \Lambda}}
\def\es{\!\!\! &=& \!\!\!}
\def\ar{&+& \!\!\!}
\def\ek{&-& \!\!\!}
\def\cp{&\times& \!\!\!}
\def\fhs{Re[FH^*]}
\def\ghs{Re[GH^*]}
\def\bcs{Re[BC^*]}
\def\fgs{Re[FG^*]}
\def\hh{|H|^2}
\def\cc{|C|^2}
\def\gg{|G|^2}
\def\ff{|F|^2}
\def\bb{|B|^2}
\def\aa{|A|^2}
\def\hh{|H|^2}
\def\ee{|E|^2}
\def\rl{\hat{m}_{\ell}}
\def\r{\hat{m}_K}
\def\s{\hat{s}}
\def\ll{\Lambda}
\def\bcds{$B_c^\ast \rightarrow D_{s}~\nu \bar{\nu} $ }
\def\bcs{$B_c^\ast $}
\def\lam{\lambda(m_{B_c^\ast}^2,m_{D_s}^2,q^2)}

\renewcommand{\textfraction}{0.2}    
\renewcommand{\topfraction}{0.8}

\renewcommand{\bottomfraction}{0.4}
\renewcommand{\floatpagefraction}{0.8}
\newcommand\mysection{\setcounter{equation}{0}\section}
\newcommand{\bra}[1]{\langle {#1}}
\newcommand{\ket}[1]{{#1} \rangle}
\newcommand{\ebar}{{\bar{e}}}
\newcommand{\sbar}{\bar{s}}
\newcommand{\cbar}{\bar{c}}
\newcommand{\bbar}{\bar{b}}
\newcommand{\qbar}{\bar{q}}
\renewcommand{\l}{\ell}
\newcommand{\lbar}{\bar{\ell}}
\newcommand{\psibar}{\bar{\psi}}
\newcommand{\barB}{\overline{B}}
\newcommand{\barK}{\overline{K}}
\newcommand{\thetaK}{\theta_{K_1}}
\newcommand{\onepone}{{1^1P_1}}
\newcommand{\sanpone}{{1^3P_1}}
\newcommand{\kone}{{K_1}}
\newcommand{\barkone}{{\overline{K}_1}}
\renewcommand{\Re}{\mathop{\mbox{Re}}}
\renewcommand{\Im}{\mathop{\mbox{Im}}}
\newcommand{\T}{{\cal T}}
\newcommand{\eff}{{\rm eff}}
\newcommand{\A}{{\cal A}}
\newcommand{\B}{{\cal B}}
\newcommand{\C}{{\cal C}}
\newcommand{\D}{{\cal D}}
\newcommand{\E}{{\cal E}}
\newcommand{\F}{{\cal F}}
\newcommand{\G}{{\cal G}}
\renewcommand{\H}{{\cal H}}
\newcommand{\hats}{\hat{s}}
\newcommand{\hatp}{\hat{p}}
\newcommand{\hatq}{\hat{q}}
\newcommand{\hatm}{\hat{m}}
\newcommand{\hatu}{\hat{u}}
\newcommand{\alphaem}{\alpha_{\rm em}}
\newcommand{\konel}{K_1(1270)}
\newcommand{\koneh}{K_1(1400)}
\newcommand{\barkonel}{\barK_1(1270)}
\newcommand{\barkoneh}{\barK_1(1400)}
\newcommand{\konea}{K_{1A}}
\newcommand{\koneb}{K_{1B}}
\newcommand{\barkonea}{\barK_{1A}}
\newcommand{\barkoneb}{\barK_{1B}}
\newcommand{\mkone}{m_{\kone}}
\newcommand{\konep}{K_1^+}
\newcommand{\konem}{K_1^-}
\newcommand{\konelm}{K_1^-(1270)}
\newcommand{\konehm}{K_1^-(1400)}
\newcommand{\konelp}{K_1^+(1270)}
\newcommand{\konehp}{K_1^+(1400)}
\newcommand{\konelz}{\overline{K}{}^0_1(1270)}
\newcommand{\konehz}{\overline{K}{}^0_1(1400)}
\newcommand{\Bm}{B^-}
\newcommand{\Bz}{\overline{B}{}^0}
\newcommand{\Kstar}{K^*(892)}
\newcommand{\BABAR}{BABAR}
\newcommand{\BELLE}{Belle}
\newcommand{\CLEO}{CLEO}
\newcommand{\leftu}{\gamma^\mu L}
\newcommand{\leftd}{\gamma_\mu L}
\newcommand{\rightu}{\gamma^\mu R}
\newcommand{\rightd}{\gamma_\mu R}
\newcommand{\Br}{{\cal B}}
\newcommand{\sect}[1]{Sec.~\ref{#1}}
\newcommand{\eqref}[1]{(\ref{#1})}
\newcommand{\fig}{FIG.~}
\newcommand{\figs}{FIGs.~}
\newcommand{\tbl}{TABLE~}
\newcommand{\tbls}{TABLEs~}
\newcommand{\errpm}[3]{#1^{+{#2}}_{-{#3}}}
\newcommand{\errpmf}[5]{{#1}^{ +{#2} +{#4} }_{-{#3}-{#5}}}
\newcommand{\lpm}{\l^+\l^-}
\newcommand{\epm}{e^+e^-}
\newcommand{\mupm}{\mu^+\mu^-}
\newcommand{\taupm}{\tau^+\tau^-}
\newcommand{\AFB}{A_{\rm FB}}
\newcommand{\barAFB}{\overline{A}_{\rm FB}}
\newcommand{\GeV}{{\,\mbox{GeV}}}
\newcommand{\MeV}{{\,\mbox{MeV}}}
\newcommand{\degree}{^\circ}
\newcommand{\mB}{m_B}
\newcommand{\SM}{{\rm SM}}
\newcommand{\NP}{{\rm NP}}
\newcommand{\barc}{\bar{c}}
\newcommand{\xipara}{\xi_\parallel^{\kone}}
\newcommand{\xiperp}{\xi_\perp^{\kone}}
\newcommand{\xiparal}{\xi_\parallel^{\konel}}
\newcommand{\xiperpl}{\xi_\perp^{\konel}}
\newcommand{\xiparah}{\xi_\parallel^{\koneh}}
\newcommand{\xiperph}{\xi_\perp^{\koneh}}
\newcommand{\para}{\parallel}
\newcommand{\alphas}{\alpha_s}
\newcommand{\pA}{p_{\kone}}
\newcommand{\lcaption}[2]{\caption{(label:{#2}) #1}\label{#2}}
\newcommand{\Rmunr}{R_{\mu,\rm nr}}
\newcommand{\RdGamma}{R_{d\Gamma/ds,\mu}}
\providecommand{\dfrac}[2]{\frac{\displaystyle
{#1}}{\displaystyle{#2}}}
\def\baeq{\begin{appeq}}     \def\eaeq{\end{appeq}}
\def\baeeq{\begin{appeeq}}   \def\eaeeq{\end{appeeq}}
\newenvironment{appeq}{\beq}{\eeq}
\newenvironment{appeeq}{\beeq}{\eeeq}
%
%
%
%
\title{\boldmath\bf
Investigation of the rare exculsive  $B_c^\ast \rightarrow D_{s}~\nu \bar{\nu}$ decays in  the faramework of the QCD sum rules}
\author{V. Bashiry\\
 Cyprus International University, Faculty of Engineering, Nicosia, Northern Cyprus, Mersin 10, Turkey\\
(e-mail: bahiry@ciu.edu.tr) }

\setlength{\baselineskip}{24pt} \maketitle
\setlength{\baselineskip}{7mm}
\date{}
\maketitle \thispagestyle{empty}
Exclusive $B_c^\ast \rightarrow D_{s}~\nu \bar{\nu}$ decay is studied in the framework of the
three--point QCD sum rules approach. The two  gluon condensate contributions to the correlation function are calculated and the form factors of this transition are found. The  decay width and total branching ratio for this decay is also calculated.
\newpage
\section{Introduction}

The standard model (SM) Higgs boson which is one of the most important components of the SM has been discovered by the ATLAS \cite{ATLAS} and CMS \cite{CMS} collaborations.  Nowadays, we aim to find  out the new physics beyond the SM.
 Heavy mesons with the different flavors like $B_c$ and $B_c^\ast$ mesons can provide a good testing benchmark not only for the predictions of the SM but also for searching the new physics beyond SM. The LHCb experiment has aimed to test  the SM predictions and discover the possible new physics signals. In this regards, a lot of the experimental data released by the LHCb experiment \cite{LHCb}.

The dominant decay mode of $B_c^\ast$ is $B_c^\ast \rightarrow B_c \gamma$~\cite{Wang:2013cha}. Rare \bcds proceeds FCNC transitions.  This decay is roughly of the same order as that of the $B_c^\ast \rightarrow \eta_c \ell \bar{\nu}_{\ell}$~\cite{ref2}. In the SM framework, the rare  \bcds decay is  dominated by the Z-penguin and box diagrams involving top quark exchanges. The theoretical uncertainties related to the renormalization scale dependence of running quark mass can be essentially neglected
after the inclusion of next-to-leading order corrections\cite{Buchalla:1998ba}. This decay is theoretically very clean processes in compare with the semileptonic decays like the $B_c^{\ast}\rightarrow D_{s}\ell^+\ell^-$  decay and is also sensitive to the new physics beyond the SM \cite{Xiao:2002ew}. Moreover, this decay is complementary to the $B_c^{\ast}\rightarrow D_{s}\ell^+\ell^-$ decay.  Note that, the direct calculation  of physical observables such as form factors suffer from sizable uncertainties. These can be greatly reduced through a
combined analysis of the rare \bcds and $B_c^{\ast}\rightarrow D_{s}\ell^+\ell^-$ \cite{Buchalla:2010jv} decays.

These decays have not yet been measured by the LHCb. There is no theoretical studies relevant to the form factors and branching ratios of \bcds decay. The form factors of these decays can be evaluated  with the different approaches. Some of them are  the light front, the constituent quark models~\cite{Geng:2001vy} and the QCD sum rules. In this study the three--point QCD sum rules approach  are used in the calculation of  form factors. It is worth mentioning that the QCD sum rules have widely  been utilized  in calculation of the   form factors (some of them can be found in Refs.\cite{Bashiry:2013waa}-\cite{Marques de Carvalho:1999ia}).

The paper has 3 sections: In section 2, the effective Hamiltonian and  the
three--point QCD sum rules approach are presented for completeness . In section 3, The numerical values of form factors are given and  the sensitivity of the branching ratio is studied and  conclusion is presented.

\section{Sum rules for the  \bcds transition form factors}\label{sec:Hamiltonian}
 The FCNC $b \rar s \nu \bar \nu$ decay is described  within the framework of the SM at the quark level by the effective Hamiltonian \cite{R13}
\bea
\label{effH}
{\cal H}_{eff} &=& \frac{G_F\alpha}{2\sqrt{2} \pi \sin^2\theta_W}
V_{tb}V_{ts}^\ast X(x) \bar b \gamma^\mu (1-\gamma_5) s \bar \nu
\gamma_\mu (1-\gamma_5) \nu~,
\eea
where $G_F$ is the Fermi  constant, $\theta_W$ is the Weinberg angle, $\alpha$ is the fine structure
coupling constant  and
\bea
\label{Xx}
X(x) = X_0(x) + \frac{\alpha_s}{4 \pi} X_1(x)~,
\eea

The $X_0(x)$  is:
\bea
\label{X0x}
X_0 = \frac{x}{8} \Bigg[\frac{x+2}{x-1} + \frac{3(x-2)}{(x-1)^2} \ln x
\Bigg]~,
\eea
where $x=m_t^2/m_W^2$.
The explicit form of  $X_1(x)$ is given in Refs.  \cite{R13} and
\cite{R14}. Note that, $X_1(x)$ gives about 3\%
contribution to the $X_0(x)$ term \cite{Aliev:2001in}.

The Wilson coefficients(in our case $X_0(x)$ and $X_1(x)$) can be calculated in any gauge and they are gauge independent and the results should be gauge invariant. The Wilson coefficients are calculated in $R_\xi$ gauge.  It is worth mentioning that  local operators in the considered problem have anomalous
dimensions.We have checked that taking into account anomalous dimensions can
change numerical results at most $10\%$.

 The matrix element  of the exclusive \bcds  decays are found by
inserting   initial meson state \bcs~  and
final meson state $D_s$~  in Eq.(\ref{effH}).
\begin{eqnarray}\label{KZVBFZ}
 M &=& \frac{G_{F}\alpha}{2\sqrt{2} \pi \sin^2\theta_W}
V_{tb}V_{ts}^{*}  X(x)
<D_{s}(p_D)\mid\overline{s} \gamma_\mu (1-\gamma_5)
b\mid B_{c}^\ast(p_B,\varepsilon)>  \overline{\nu}\gamma_\mu (1-\gamma_5) \nu
\end{eqnarray}
where $\varepsilon$ is the polarization vector of \bcs~  meson, $p_B$ is the  momentum  of the $B_c^\ast$~ and $p_D$ is the momentum of  $D_s$~ meson. The  matrix element of the Eq. (\ref{KZVBFZ}) is
written in terms of the form factors as follows:
\begin{eqnarray}\label{3au}
\nonumber <D_{s}(p_D)\mid\overline{s}\gamma_{\mu}(1-\gamma_{5}) b\mid
B_c^\ast(p_B,\varepsilon)>&=&\frac{A_{V}(q^2)}{m_{B_c^\ast}}\varepsilon_{\mu\nu\alpha\beta}
\varepsilon^{\ast\nu}p_B^\alpha p_D^\beta-iA_{0}(q^2)m_{B_c^\ast}\varepsilon_{\mu}^{\ast} \\
-i\frac{A_{+}(q^2)}{m_{B_c^\ast}}(\varepsilon^{*}p_D)P_{\mu}
&-&i \frac{A_{-}(q^2)}{m_{B_c^\ast}}(\varepsilon^{*}p_D)q_{\mu},
\end{eqnarray}
here, Lorentz invariant and parity conservation are considered. Also, $A_{i}(q^2)$, where $i=V,+ , -$   are the dimensionless  transition form factors. $P_{\mu}=(p_B+p_D)_{\mu}$ and $q_{\mu}=(p_B-p_D)_{\mu}$ is the transfer momentum or the momentum of the $Z$ boson.

The matrix element in terms of the form factors is as:

\begin{eqnarray}\label{ampilitude}
 M &=& \frac{G_{F}\alpha}{2\sqrt{2} \pi \sin^2\theta_W}
V_{tb}V_{ts}^{*}  X(x)\Big[i \frac{A_{1}(q^2)}{m_{B_c^\ast}}\varepsilon_{\mu\nu\alpha\beta}
\varepsilon^{\ast\nu}p_B^\alpha p_D^\beta-iA_{0}(q^2)m_{B_c^\ast}\varepsilon_{\mu}^{\ast}
\nonumber\\ &-&i\frac{A_{+}(q^2)}{m_{B_c^\ast}}(\varepsilon^{*}p_D)P_{\mu}
-i \frac{A_{-}(q^2)}{m_{B_c^\ast}}(\varepsilon^{*}p_D)q_{\mu}\Bigg]  \overline{\nu}\gamma_\mu (1-\gamma_5) \nu  ,
\end{eqnarray}

where$A_1=-i A_V$

 We try to calculate the  the aforementioned form factors by means of the QCD sum rules.  The QCD sum
rules begin with the  the following correlation functions:
\begin{eqnarray}\label{6au}
\Pi _{\mu\nu}^{V-AV}(p_B^2,p_D^2,q^2)&=&i^2\int
d^{4}xd^4ye^{-ip_Bx}e^{ip_Dy}<0\mid T[J_{D_{s}}(y)
J_{\mu}^{V-AV}(0) J_{\nu B_{c}^\ast}(x)]\mid  0>,
\end{eqnarray}
where the interpolating currents are  $J _{ D_{s}}(y)=\overline{c}\gamma_{5} s$ and
$J_{\nu B_{c}^\ast}(x)=\overline{b}\gamma_{\nu}c$   the   $D_s$ and   the \bcs~ meson states, respectively.
 $J_{\mu}^{V-AV}=~\overline {s}\gamma_{\mu}(1-\gamma_{5})b $  consists of the  vector ($V$) and axial vector ($AV$) transition currents.
After inserting the  the two complete sets of the \bcs ~and $D_s$ meson,
the correlation functions in Eq.~(\ref{6au})  is written  as follows:

\begin{eqnarray} \label{7au}
&&\Pi _{\mu\nu}^{V-AV}(p_B^2,p_D^2,q^2)=-
\nonumber \\
&&\frac{<0\mid J_{D_{s}} \mid
D_{s}(p_D)><D_{s}(p_D)\mid
J_{\mu}^{V-AV}\mid B_{c}^\ast(p_B,{\varepsilon})><B_{c}^\ast(p_B,{\varepsilon})\mid J_{\nu B_c^\ast}\mid
0>}{(p_D^2-m_{D_{s}}^2)(p_B^2-m_{B_c^\ast}^2)}+\cdots,
\end{eqnarray}
where "$\cdots$" shows the contributions come from  higher states and continuum of the currents with the same quantum numbers.

The $<0\mid J_{D_{s}} \mid D_{s}(p_D)>$ and $<B_{c}^\ast(p_B,{\varepsilon})\mid J_{\nu B_c^\ast}\mid 0>$ matrix  elements are defined  as follows:
\begin{equation}\label{8au}
 <0\mid J_{D_{s}} \mid
D_{s}(p_D)>=-i\frac{f_{D_{s}}m_{D_{s}}^2}{m_{s}+m_{c}} ~,~~<B_{c}^\ast(p_B,{\varepsilon})\mid J_{\nu B_c^\ast}\mid
0>=f_{B_{c}^\ast}m_{B_{c}^\ast}\varepsilon_\nu,
\end{equation}

where $f_{B_{c}}$ and $f_{D_s}$  are the leptonic decay
constants of $B_{c}^\ast$ and $D_{s} $ mesons, respectively. Using
theses equations and calculating the   the
summation over the polarization of the vector meson $B_c^\ast$ , the
Eq.(\ref{7au}) is as follows:
\begin{eqnarray}\label{9amplitude}
\Pi_{\mu\nu}^{V-AV}(p_B^2,p_D^2,q^2)&=&-\frac{f_{D_{s}}m_{D_{s}}^2}
{(m_{c}+m_{s})}\frac{f_{B_{c}^{\ast}}m_{B_{c}^\ast}}
{(p_D^2-m_{D_{s}}^2)(p_B^2-m_{B_c^\ast}^2)} \times
\left[\vphantom{\int_0^{x_2}}A_{0}(q^2)m_{B_c^\ast}g_{\mu\nu}+\frac{A_{+}(q^2)}{m_{B_c^\ast}}P_{\mu}p_{B\nu}\right.
\nonumber
\\ &+&\left. \frac{A_{-}(q^2)}{m_{B_c^\ast}}q_{\mu}p_{B\nu}+i\frac{A_{1}(q^2)}{m_{B_c^\ast}}\varepsilon_{\mu\nu\alpha\beta}
p_B^{\alpha}p_D^{\beta}\vphantom{\int_0^{x_2}}\right] + \mbox{excited states,}
\end{eqnarray}
This  correlation function is calculated in terms of the  quarks and gluons parameters by means of the  the operator product expansion (OPE) as:
\begin{eqnarray}\label{QCD side}
\Pi_{\mu\nu}^{V-AV}(p_B^2,p_D^2,q^2)&=&\Pi^{V-AV}_{0}m_{B_c^\ast}g_{\mu\nu}
+\frac{\Pi^{V-AV}_{+}}{m_{B_c^\ast}}P_{\mu}p_{B\nu}+\frac{\Pi^{V-AV}_{-}}{m_{B_c^\ast}}q_{\mu}p_{B\nu}+
i\frac{\Pi^{V-AV}_{1}}{m_{B_c^\ast}}\varepsilon_{\mu\nu\alpha\beta}p_B^{\alpha}p_D^{\beta},
\end{eqnarray}
 Each $\Pi_{i}$ with $i=0, +, -$ and $1$ contains of the perturbative and non-perturbative parts as in the following:
\begin{eqnarray}\label{QCD side1}
\Pi_{i}&=&\Pi_{i}^{pert}+ \Pi_{i}^{nonpert}.
\end{eqnarray}
The bare-loop diagram given in Fig.1(a) is the contribution of the perturbative part.  The non-perturbative part consists of the two gluon condensates diagrams \{see Fig.2(a-f)\}. Hence, contributions of
the light quark condensates \{diagrams shown in Fig.1(b, c, d)\} vanish  by applying the
double Borel transformations \cite{Azizi:2008vv}.
\begin{figure}
  \centering
  \includegraphics[width=11cm]{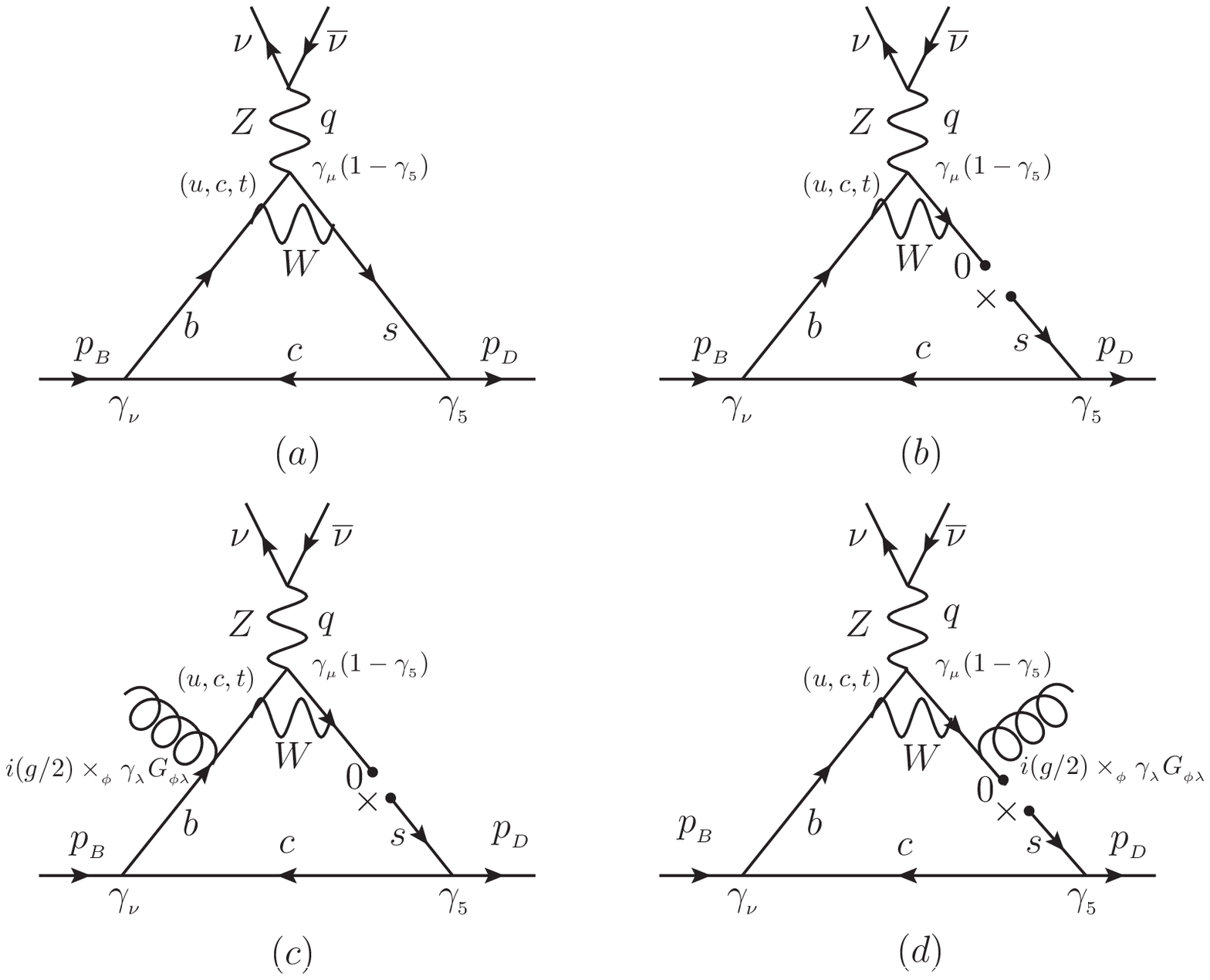}\\
  \caption{The bare-loop and light quarks condensates contributions to $B_c^* \rightarrow
D_{s}~l^+ l^-$ transitions}\label{fig1}
\end{figure}

 The following  double dispersion integrals are the contributions of the bare-loop diagrams  in the correlation function:

\begin{equation}\label{10au}
\Pi_i^{per}=-\frac{1}{(2\pi)^2}\int du\int
ds\frac{\rho_{i}(s,u,q^2)}{(s-p_B^2)(u-p_D^2)}+\textrm{ subtraction
terms}.
\end{equation}
One of the basic methods to solve the Feynman Integrals in order to calculate the spectral densities $\rho_{i}(s,u,q^2)$ is  Cutkosky rules where the quark propagators are replaced by Dirac Delta Functions:
$\frac{1}{p^2-m^2}\rightarrow-2\pi i\delta(p^2-m^2),$ which indicates that all quarks are on-shell.

Three delta functions appear as a result of the applying Cutkosky rules. These delta functions have to vanish at the same time. Therefore, we get the  following inequality from the arguments of the delta functions:
 \begin{equation}\label{13au}
 -1\leq\frac{2su+(s+u-q^2)(m_{b}^2-s-m_{c}^2)+(m_{c}^2-m_{s}^2)2s}
 {\lambda^{1/2}(m_{b}^2,s,m_{c}^2)\lambda^{1/2}(s,u,q^2)}\leq+1
\end{equation}
where $ \lambda(a,b,c)=a^2+b^2+c^2-2ac-2bc-2ab$.

Following the standard  calculations, the spectral densities are evaluated as:

\bea \rho^{V-AV}_1&=&N_c I_0(s,u,q^2)\Bigg\{C_1
(m_b-m_c)-(C_2+1) m_c+C_2 m_s\Bigg\}\nnb\\
 \rho^{V-AV}_0 &=&\frac{N_c}{2} I_0(s,u,q^2)\Bigg\{-2
m_c^3+2 m_s m_c^2-[(C_1+C_2+1) (-q^2+s+u)+2 C_1 s
 \nnb \\ &+&
2 C_2 u] m_c+ m_b[2 m_c^2-2 m_s m_c+2 C_2u+C_1 (-q^2+s+u)]+m_s
[2 C_1 s
\nnb\\ &+& C_2 (-q^2+s+u)] \Bigg\}\nnb\\
 \rho^{V-AV}_{+}&=&\frac{N_c}{2} I_0(s,u,q^2)\Bigg\{
C_1(m_b-2C_2m_c-m_c+2C_2 m_s)\nnb\\&-&
(2 C_2+1) (C_2m_c+m_c-C_2 m_s) \Bigg\}\nnb\\
 \rho^{V-AV}_{-}&=&\frac{N_c}{2} I_0(s,u,q^2)\Bigg\{(2C_2-1) (C_2 m_c+m_c-C_2 m_s)
\nnb\\&+&C_1 (m_b-2 C_2m_c-m_c+2 C_2 m_s) \Bigg\}\nnb\\
\eea

where
\bea
I_{0}(s,u,q^2)&=&\frac{1}{4\lambda^{1/2}(s,u,q^2)},\nonumber\\
C_1&=&\frac{m_c^2 (s - u - q^2) + u (2 m_b^2 - s + u - q^2) - m_s^2 (s + u - q^2)}{\lambda(s,u,q^2)}\nnb\\
C_2&=&\frac{s (2 m_s^2 + s - u - q^2) - m_b^2 (s + u - q^2) - m_c^2 (s - u + q^2)}{\lambda(s,u,q^2)}\nnb\\
N_c&=&3.
\eea
Now, it is aimed to calculate  the non-perturbative part of the Eq.(\ref{QCD side1}) which consists of the gluon condensates diagrams shown in Fig.2. The gluon condensate contributions are calculated in Fock-Schwinger gauge
because in this gauge the gluon field is expressed in terms of gluon field strength tensor directly.
\begin{figure}
\vspace*{-1cm}
\begin{center}
\includegraphics[width=11cm]{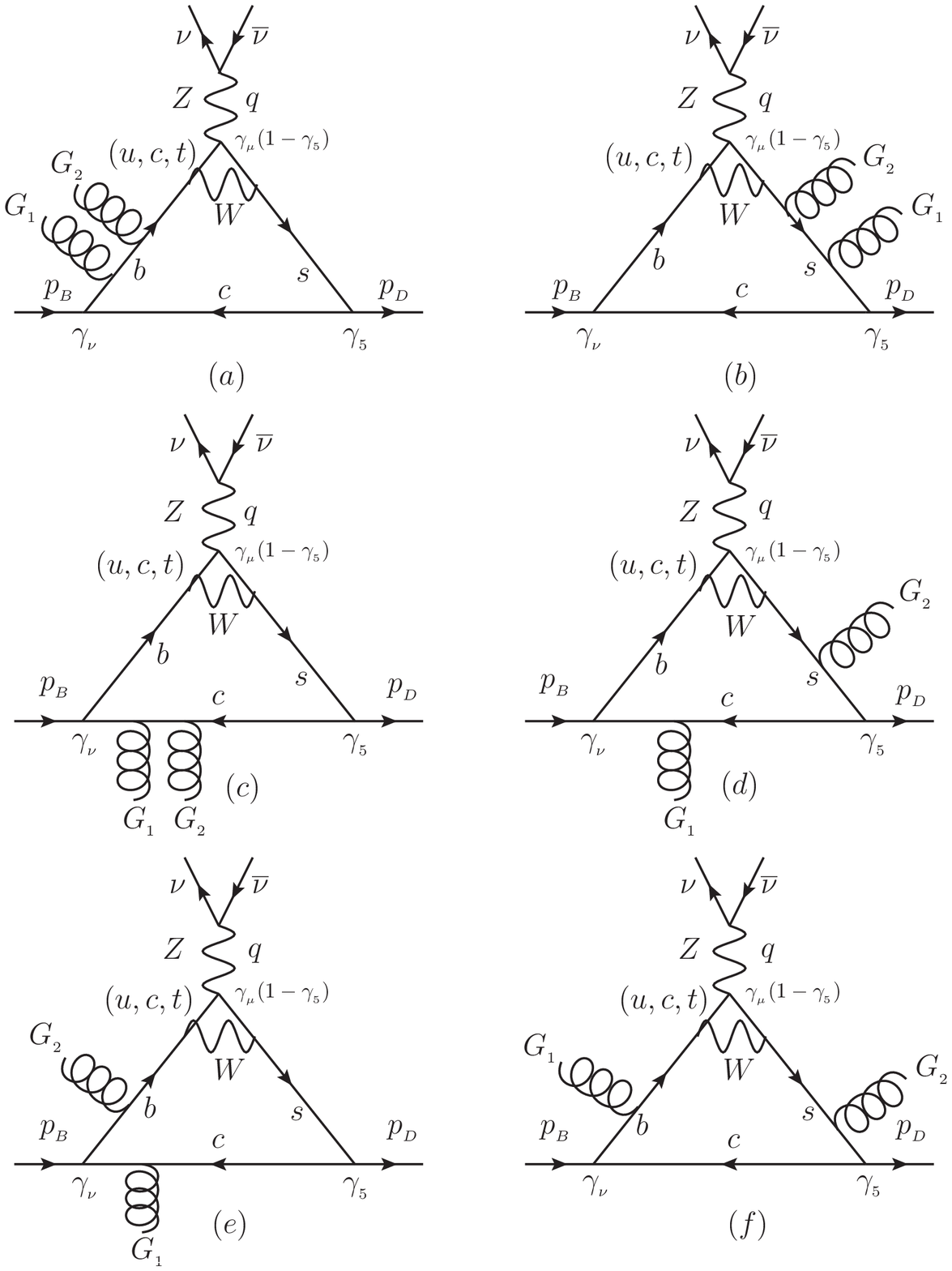}
\end{center}
\caption{Gluon condensate contributions to $B_c^* \rightarrow
D_{s}~\nu^+ \nu^-$ transitions  } \label{fig2}
\end{figure}
The following  type of the integrals has to  be calculated in order to get the results of the   gluon condensate diagrams \cite{Aliev:2006vs,R7323}: \bea \label{e7323} I_0[a,b,c] = \int
\frac{d^4k}{(2 \pi)^4} \frac{1}{\left[ k^2-m_b^2 \right]^a \left[
(p_B+k)^2-m_c^2 \right]^b \left[ (p_D+k)^2-m_s^2\right]^c}~,
\nnb \\ \nnb \\
I_\mu[a,b,c] = \int \frac{d^4k}{(2 \pi)^4} \frac{k_\mu}{\left[
k^2-m_b^2 \right]^a \left[ (p_B+k)^2-m_c^2 \right]^b \left[
(p_D+k)^2-m_s^2\right]^c}~,
\nnb \\ \nnb \\
I_{\mu\nu}[a,b,c] =\int \frac{d^4k}{(2 \pi)^4} \frac{k_\mu
k_\nu}{\left[ k^2-m_b^2 \right]^a \left[ (p_B+k)^2-m_c^2 \right]^b
\left[ (p_D+k)^2-m_s^2\right]^c}~, \eea where $k$ is the
momentum of the spectator quark $c$.
The generic solutions for these integrals can be seen in Refs. \cite{R7323}-\cite{R7321}.
Apart of our results for the contributions of the gluon condensate diagrams following the similar methods shown in Refs.\cite{R7323}- \cite{R7321} is given in Appendix.

The Borel transformations are  applied for both phenomenological and QCD side \{Eq.~(\ref{QCD side})\} in order to  suppress the contributions of higher states and continuum. The QCD sum rules for the form factors ( $A_{V}$, $A_{0}$, $A_{+}$ , and $A_{-}$  are  obtained
by equalizing   the  Borel transformed forms  of the physical  side. The result is in the following formula:
\begin{eqnarray}\label{15au}
A_{i}(q^2)&=&\frac{(m_{s}+m_{c})e^{m_{B_{c}^*}^2/M_{1}^2}e^{m_{D_s}^2/M_{2}^2}
}{f_{B_{c}^*}m_{B_{c}^*}f_{D_s}m_{D_s}^2}\left[\vphantom{\int_0^{x_2}}\frac{1}{(2\pi)^2}\int_{u_{min}}^{u_0}
du
 \int_{s_{min}}^{s_0} ds\rho_{i}^{V-AV}(s,u,q^2)e^{-s/M_{1}^2-u/M_{2}^2}\right.\nonumber
\\&+&\left.i\frac{1}{24\pi^2}{C^{A_{i}}}<\frac{\alpha_{s}}{\pi}G^{2}>\vphantom{\int_0^{x_2}}\right]
 \end{eqnarray}
Note that, the contributions of the gluon condensates ($C^{A_{i}}$ ) are already considered in the numerical analysis. However,  each of these explicit expressions are extremely long, it is found unnecessary to show all of them  in this study. Therefore, one of these expressions ($C^{A_{V}}$) is shown as a sample in Appendix.
The $s_{0}$ and $u_{0}$ are the continuum
thresholds in $s$ and $u$ channels, respectively. Also $s_{min}=(m_b+m_c)^2$ and $u_{min}=(m_s+m_c)^2$.

\section{Numerical analysis}
Having known the matrix element i.e., Eq. (\ref{ampilitude}),  the decay rate for \bcds  decay is evaluated  as follows:
\bea\label{decay} \frac{d\Gamma}{dq^2}&=&\frac {\alpha ^2 G_f^2 \lambda^{1/2}(m_{B_c^\ast}^2,m_{D_s}^2,q^2) |V_{tb}V_{ts}^\ast|^2v X^2(x)}{3072 \pi ^5
      {m^3_ {B_c^\ast}} \sin^4\theta_W}
     \bigg\{  |A_ 0|^2(m^4_{B_c^\ast}-2 m^2_{B_c^\ast}(m_{D_s}^2-5q^2)+(m_{D_s}^2-q^2)^2)\nnb\\&-&2\Re[A_+A_0^\ast]\frac{m^6_{B_c^\ast}
     -(m_{D_s}^2-q^2)^3-m^4_{B_c^\ast}(3m_{D_s}^2+q^2)+m^2_{B_c^\ast}(3m_{D_s}^4-2 m^2_{D_s}q^2-q^4)}{m_{B_c^\ast}^2}\nnb\\&+&2|A_1|^2q^2\frac{\lam}{m_{B_c^\ast}^2}+
    |A_+|^2 \frac{\lambda^{2}(m_{B_c^\ast}^2,m_{D_s}^2,q^2)}{m^4_{B_c^\ast}}\bigg\} \eea

The expression for  the decay rate shows that we need to know  the input parameters shown in table \ref{input}, taken from Ref.\cite{pdg12}.
\begin{table}[h]
\center
\begin{tabular}{|c|c|}
  \hline
 $\mid V_{tb}\mid$ &$0.77^{+0.18}_{-0.24}$ \\ \hline
  $\mid V_{ts}\mid $& $(40.6\pm2.7)\times10^{-3}$ \\ \hline
 $\tau_{B_c^\ast}$ & $(0.452 \pm0.033) \times 10^{-12} s$ \\ \hline
 $\alpha(m^2_w)$ &$1/128$  \\ \hline
  $\sin^2\theta_W$ &  $0.2315$ \\ \hline
 $m_t$ & $173.07 \pm0.52 \pm0.72$ GeV \\ \hline
  $m_W $& $80.385 \pm0.015$ GeV \\ \hline
   $m_{B_c^\ast}$& $6.2745\pm 0.0018$ GeV \\ \hline
    $m_{D_s}$& $1968.50\pm0.32 $ Mev \\ \hline
     $f_{D_{s}}$& $(206.7 \pm 8.5 \pm2.5)$ MeV\\ \hline
      $m_{b} $& $(4.18\pm 0.03)~GeV$  \\ \hline
       $ m_{c}(\mu=m_{c}) $& $1.275\pm 0.015~ GeV$ \\ \hline
   \end{tabular}
\caption{The values of the input parameters\cite {pdg12} }\label{input}
\end{table}

 Moreover, the values of the leptonic decay constants $f_{B_{C}^{\ast}}=0.415\pm0.031$GeV\cite{Wang:2012kw} and the gluon condensate $<\frac{\alpha_{s}}{\pi}G^{2}>=0.012~ GeV ^{4}$
\cite{Shifman1} are necessary for the evaluation of the form factors.
In addition, the form factors  contain four auxiliary
parameters: the Borel mass squares $M_{1}^2$ and $M_{2}^2$ and the continuum
threshold $s_{0}$  and $u_{0}$. The  form factors are assumed to be independent or  weakly dependent  on
these auxiliary parameters in the  suitable chosen regions named as  "working regions".

The contributions proportional to the highest
power of $1/M_{1,2}^{2}$ are supposed to be  less than about $30^{0}/_{0}$ of  the contributions proportional to the
highest power of $M_{1,2}^{2}$. The lower bound of the  $M_{1}^2$ and $M_{2}^2$ can be determined by the above condition. In addition, the contributions of continuum must be  less than that of the first resonance. This helps us to fix the upper bound of the $M_{1}^2$ and $M_{2}^2$. Therefore, we find the suitable  region for the Borel  mass parameters in the following intervals; $10~GeV^2\leq M_{1}^{2}\leq25~GeV^2 $ and $4~GeV^2\leq M_{2}^{2}\leq10~GeV^2 $.

The numerical value of the $s_0$ and $u_0$ are supposed to be less than the
mass squared  of the first excited state meson with the same quantum numbers. In other words, the $s_0$ and $u_0$ are between mass squared of the ground sate meson and excited state meson with the same quantum numbers. The following regions for the  $s_0$ and $u_0$ are chosen:
$(m_{B_c^\ast}+0.3)^2\leq s_0\leq(m_{B_c^\ast}+0.7)^2$  and $(m_{D_s}+0.3)^2\leq u_0\leq (m_{D_s}+0.7)^2$ .

The form factors depend on the $q^2$. The detail of the dependence is complicated. We fit them to the  following function:

\begin{eqnarray}\label{fitfunction}
F(q^2)=\frac{a}{1- q^2/m_{fit}^2}+\frac{b}{(1-q^2/m_{fit}^2)^2}
\end{eqnarray}

The $a$ , $b$ and $m_{fit}$ are given in Table \ref{tabpi}:
\begin{table}[h]
\center
\begin{tabular}{|c|c|c|c|}
\hline
& $m_{fit}$ & $a$& $b$\\
\hline
$A_1(q^2)$ & $5.01\pm 1.1 $ & $ -0.14\pm 0.04$ &$ 0.26\pm 0.08$ \\
\hline
$A_0(q^2)$ & $6.44\pm 1.4 $ & $ -0.11\pm0.03$ &$ 0.17\pm0.06$ \\
\hline
$A_+(q^2)$ & $5.00\pm 1.08 $ & $ -0.14\pm 0.04$ &$ 0.28\pm 0.08$ \\
\hline
$A_-(q^2)$ & $4.98\pm 1.07 $ & $ -0.14\pm 0.04$ &$ 0.28\pm 0.08$  \\
\hline
\end{tabular}
\caption{Parameters appearing in the form factors of the
\bcds decay in a four-parameter fit, for $M_{1}^2=15~GeV^2$, $M^2_2= 6$ GeV$^2$, $s_0=46$GeV$^2$ and $u_0=6$GeV$^2$}\label{tabpi}
\end{table}

The origin of the errors  in Table II are the variation of $s_0$, $u_0$ and $M_{1,2}$  in the chosen intervals and the uncertainties of the input parameters.

\begin{figure}
  \centering
  \includegraphics[width=11cm]{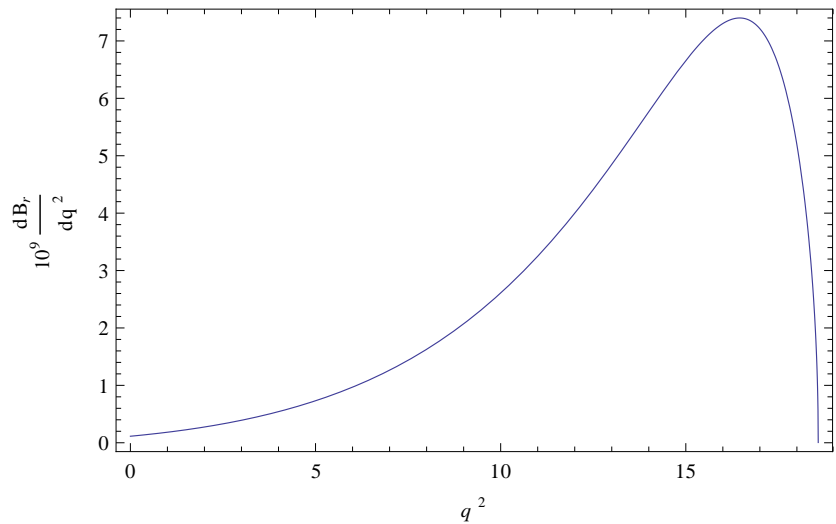}\\
  \caption{The dependence  of the branching ratio on $q^2$  for $B_c^* \rightarrow
D_{s}~\mu^+ \mu^-$ transitions}\label{fig3}
\end{figure}

In order to evaluate the branching ratio of the \bcds decay, the mean life time of the $B_c^\ast$ meson is needed. For the time being there is no experimental data on the mean life time of this meson. We follow the theoretical methods like  Bethe-Salpeter model \cite{AbdElHady:1998kc} and potential model \cite{Kiselev:2000jc}, and estimate that the mean life time of the $B_c^\ast$ meson is in the order of the mean life time of the $B_c$ meson. We assume that
 the total life-time $\tau_{B_{c}}\approx\tau_{B_{c}}=0.452 \times10^{-12}s$ \cite{pdg12}. Using the mean life time and the $q^2$ dependence of the form factors given by Eq.(\ref{fitfunction}) in the kinematical allowed region$[0\leq q^2\leq (m_{B_c^\ast}-m_{D_{s}} )^{2}]$ we study the
branching ratios for \bcds ~decay. Our results for three different values of the $q^2=(1,6,12)$~GeV$^2$ are presented in Tables \ref{branch}. In addition, Fig. (\ref{fig3}) depicts the dependence of the branching ratio on $q^2$ for full kinematical allowed region.

\begin{table}[h]
\centering
\begin{tabular}{|c|c|}
 \hline
  $q^2($GeV$^2)$ & ${\cal B}_r(q^2)(B_{c}^\ast\rightarrow D_{s}\nu\bar{\nu}$) \\
  \hline
$1$& $1.83\times 10^{-10}$\\
\hline
$6$&$9.68\times 10^{-10}$   \\
\hline
 $12$&$3.99\times 10^{-9}$\\
 \hline
  \end{tabular}
\caption{Values for the branching ratio of  the  \bcds decay at three different values of the dileptonic invariant mass.} \label{branch}
\end{table}

Finally, we calculate   the integrated  branching
ratio for \bcds decay  as follows:
\begin{equation}\label{intbr}
  {\cal B}_r=\int^{(m_{B_c^\ast}-m_{D_{s}} )^{2}}_{0} {\cal B}_r(q^2)dq^2=(5.47\pm 1.30)\times10^{-8}
\end{equation}

To sum up, we investigated the branching ratio and decay rate of the \bcds
decay.  The form factors of this decay were found in the framework of the QCD sum rules.
In addition,  the contributions of the two  gluon condensates diagrams to the correlations function were obtained.

\newpage


\newpage

\section*{Appendix}
In this section, we present the explicit expression for the
coefficients $C^{A_{V}}$ corresponding to the
gluon condensates contributions of $g_{\mu\nu}$ structure entering to the expression for the form factors
in Eq.(\ref{15au}).
\bea
C^{A_{V}}&=&(8 m_b+16 m_c) I(1,1,2)-(32 m_c^3+16 m_b m_c^2+8 m_c
q^2) I(1,1,3)
   -(16 m_c^3+8 m_b m_c^2
 \\ \nnb &+&
   +8 m_c q^2) I(1,2,2)+(8 m_c^5-16 m_b m_c^4-8 m_c^3 q^2) I(1,2,3)+(-24 m_b m_c^2-24 m_b^2 m_c)I(1,3,1)
      \\ \nnb &-&
   24 m_b m_c^4 I(1,3,2)-8 m_b m_c^6I(1,3,3)+8 m_c I(2,1,1)+(-16m_b^3-8 m_c m_b^2-8 q^2 m_b) I(2,2,1)
   \\ \nnb &-&
   24 m_b^4 m_c) I(2,3,1)+(32 m_b^3-16 m_c m_b^2-8 q^2 m_b) I(3,1,1)-
   (8 m_b^5+16 m_c m_b^4+8 q^2 m_b^3) I(3,2,1)
   \\ \nnb &-&
   8m_b^6 m_c I(3,3,1)+(8 m_b q^2-8 m_c q^2)I_1(1,1,3)-24 m_b m_c^2 q^2 I_1(1,1,4)+8 m_b q^2I_1(1,3,1)
   \\ \nnb &+&
   (8 m_b q^2-8 m_c q^2)I_1(2,1,2)+(-16 m_b q^4+8 m_c q^4-24 m_c^3 q^2+24 m_b m_c^2
   q^2+8 m_b^2 m_c q^2)I_1(2,1,3)
   \\ \nnb &+&
   16 q^2 m_b^3 I_1(2,3,1)
     -16 m_b q^2 I_1(3,1,1)+(-16 m_b q^4+24 m_b^3 q^2+16 m_b m_c^2 q^2)I_1(3,1,2)
   \\ \nnb &+&
   (8 m_b q^6-16 m_b^3 q^4-16 m_b m_c^2 q^4+8 m_b^5 q^2+8 m_b m_c^4 q^2-16 m_b^3m_c^2 q^2)
   I_1(3,1,3)+16 q^2 m_b^3I_1(3,2,1)
   \\ \nnb &+&
   8 q^2 m_b^5 I_1(3,3,1)+72 m_b^3 q^2 I_1(4,1,1)-16 m_c q^2 I_2(1,1,3)-24 m_c^3 q^2 I_2(1,1,4)-8m_c q^2 I_2(1,2,2)
   \\ \nnb &+&
   16 m_c^3 q^2I_2(1,2,3)+8 m_c q^2 I_2(1,3,1)+16 m_c^3 q^2 I_2(1,3,2)+8 m_c^5 q^2 I_2(1,3,3)
   \\ \nnb &+&
   (8 m_c q^2-8 m_b q^2) I_2(2,1,2)+(40 m_c^3 q^2-16 m_c q^4)I_2(2,1,3)+(-8 m_b q^2-40 m_c q^2) I_2(3,1,1)
   \\ \nnb &+&
   (8 m_b q^4-16 m_c q^4-8 m_b^3q^2+16 m_c^3 q^2-8 m_b m_c^2 q^2+8 m_b^2 m_c q^2)
   I_2(3,1,2)+(8 m_c q^6-16 m_c^3 q^4
   \\ \nnb &-&
   16m_b^2 m_c q^4+8 m_c^5 q^2-16 m_b^2 m_c^3 q^2+8 m_b^4 m_c q^2)
   I_2(3,1,3)+72 m_b^2 m_c q^2 I_2(4,1,1)
   \nnb\\ \nnb&+&
   D_3^0 \Bigg\{(8 m_c-8 m_b)I_1(3,3,1)\Bigg\}+D^3_0 \Bigg\{8 m_b I(1,3,3)+8m_c I_2(1,3,3)\Bigg\}
\\ \nnb &+&
D_0^2 \Bigg\{(-24 m_b+8 m_c) I(1,2,3)+(8 m_c-24 m_b)I(1,3,2)+(8
m_c^3-24 m_b m_c^2-8 q^2 m_c)I(1,3,3)
\\ \nnb &-&
   16 m_c I_2(1,2,3)-8 m_c I_2(1,3,2)-(16 m_c^3+8 q^2 m_c) I_2(1,3,3)\Bigg\}
   \\ \nnb &+&
   D^0_2\Bigg\{ D^1_0\Bigg[8 m_c I(3,3,1)+(8 m_c-8 m_b) I_1(3,3,1)+(16 m_c-16 m_b)
   I_2(3,3,1)\Bigg]+8m_c I(2,3,1)
\\ \nnb &+&
   (-16 m_b+8 m_c) I(3,2,1)+(8 m_c^3-16 m_b m_c^2+ 8 m_b^2 m_c-8 q^2 m_c) I(3,3,1)
   \\ \nnb &+&
   (16m_b-8 m_c) I_1(2,3,1)+(8 m_b-16 m_c) I_1(3,2,1)
   \\ \nnb &+&
   (16 m_b^3-16 m_c m_b^2+16 m_c^2 m_b+8 q^2 m_b-16 m_c^3-8 m_c q^2) I_1(3,3,1)\Bigg\}
   \\ \nnb &+&
   D_0^1 \Bigg\{ D^2_0\Bigg[8 m_c I(1,3,3)+16 m_c I_1(1,3,3)+8 m_c
   I_2(1,3,3)\Bigg]+D^1_0\Bigg[-16 m_c I(1,2,3)
   \\ \nnb &-&
   16 m_c I(1,3,2)-16 m_c^3 I(1,3,3)-16 m_c I(2,3,1)+(-8 m_b-16 m_c) I(3,2,1)
   \\ \nnb &+&
   (-16 m_c^3-16 m_b^2 m_c) I(3,3,1)-32 m_c I_1(1,2,3)-16 m_c I_1(1,3,2)-32 m_c^3 I_1(1,3,3)
   \\ \nnb &+&
   (16 m_b-8 m_c) I_1(2,3,1)+(8 m_b-16 m_c) I_1(3,2,1)+(16 m_b^3-16 m_c m_b^2+16 m_c^2 m_b
   \\ \nnb &-&
   16 m_c^3) I_1(3,3,1)-16 m_c I_2(1,2,3)-8 m_c I_2(1,3,2)-16
   m_c^3I_2(1,3,3)+(32 m_b-16 m_c) I_2(2,3,1)
   \\ \nnb &+&
   (16 m_b-32 m_c) I_2(3,2,1)+(32 m_b^3-32 m_c m_b^2+32 m_c^2 m_b-32 m_c^3) I_2(3,3,1)\Bigg]
   \\ \nnb &+&
    8 m_c I(1,1,3)+16 m_c^3 I(1,2,3)-32 m_c I(1,3,1)+24 m_c^3 I(1,3,2)+8 m_c^5 I(1,3,3)-24 m_c^3 I(1,4,1)
\\ \nnb &+&
    (24 m_b-16 m_c) I(2,2,1)+(-16 m_c^3+ 32 m_b m_c^2-32 m_b^2 m_c+16 q^2 m_c) I(2,3,1)
   \\ \nnb &+&
   (24 m_b-64 m_c) I(3,1,1)+(24 m_b^3-32 m_c m_b^2+56 m_c^2 m_b+8 q^2 m_b-32 m_c^3+16 m_c q^2) I(3,2,1)
\\ \nnb &+&
   (-16 m_c^5 +32 m_b m_c^4-32 m_b^2 m_c^3+16 q^2 m_c^3+32 m_b^3
   m_c^2-16m_b^4 m_c+16 m_b^2 q^2 m_c) I(3,3,1)
 \\ \nnb &+&
   72 m_b^2 m_c I(4,1,1)+(-8 m_b+16 m_c) I_1(1,1,3) -16 m_c I_1(1,2,2)+32 m_c^3 I_1(1,2,3)
   \\ \nnb &+&
   (-8 m_b-72 m_c) I_1(1,3,1)+32 m_c^3 I_1(1,3,2)+16 m_c^5 I_1(1,3,3)+(-72 m_c^3+24 m_b m_c^2) I_1(1,4,1)
   \\ \nnb &+&
   8 m_b I_1(2,1,2)+16 m_b q^2 I_1(2,1,3)+(-16 m_b^3+8 m_c m_b^2-8 m_c^2 m_b-16 q^2 m_b+16 m_c^3
   \\ \nnb &+&
   8 m_c q^2) I_1(2,3,1)+(32 m_b-136 m_c) I_1(3,1,1)-8 m_b^3 I_1(3,1,2)+(-8 m_b^5+16 q^2 m_b^3-8 q^4 m_b) I_1(3,1,3)
   \\ \nnb &+&
   (-16 m_b^3+8 m_c m_b^2-8 m_c^2 m_b-8 q^2 m_b+16 m_c^3+16 m_c q^2) I_1(3,2,1)+(-8
   m_b^5+8 m_c m_b^4+16 m_c^2 m_b^3
   \\ \nnb &-&
   16 q^2 m_b^3-16 m_c^3 m_b^2+16 m_c q^2 m_b^2-8 m_c^4 m_b-16 m_c^2 q^2
   m_b+8m_c^5+16 m_c^3 q^2) I_1(3,3,1)
\\ \nnb &+&
   (-72 m_b^3+216 m_c m_b^2)I_1(4,1,1)+
   8 m_c I_2(1,1,3)-8m_c I_2(1,2,2)+16 m_c^3 I_2(1,2,3)-24m_c I_2(1,3,1)
   \\ \nnb &+&
   16 m_c^3I_2(1,3,2)+8 m_c^5 I_2(1,3,3)-24 m_c^3 I_2(1,4,1)+8 m_bI_2(2,1,2)+(8 m_b-48 m_c) I_2(3,1,1)
\\ \nnb &+&
   (8m_b^3-8 q^2 m_b) I_2(3,1,2)+72 m_b^2 m_c I_2(4,1,1)\Bigg\}
   \\ \nnb &+&
   D^1_0 \Bigg\{-24m_c^3 I(1,4,1)  +72 m_b^2 I(4,1,1) m_c+(24 m_b-16 m_c) I(1,1,3)+(16 m_b-16 m_c) I(1,2,2)
\\ \nnb &+&
   (-32 m_c^3+48 m_b m_c^2+16 q^2 m_c) I(1,2,3)+ (24 m_b-32 m_c) I(1,3,1)+(-16 m_c^3+40 m_b m_c^2
\\ \nnb &+&
  16 q^2 m_c) I(1,3,2)+(-16 m_c^5+24 m_b m_c^4+16 q^2 m_c^3) I(1,3,3)+(8 m_b+8 m_c) I(2,2,1)
   \\ \nnb &+&
   (24 m_c^3+16 m_b^2 m_c) I(2,3,1)+(8m_b-40 m_c) I(3,1,1)+(8 m_b^3+8 m_c m_b^2+8 m_c^2
   m_b+16 m_c^3)I(3,2,1)
   \\ \nnb &+&
   (8 m_c^5-16 m_b^2 m_c^3+8 m_b^4 m_c) I(3,3,1)-8 m_bI_1(1,1,3)+(-8 m_b-24 m_c)I_1(1,3,1)+(24 m_b m_c^2
   \\ \nnb &-&
   24 m_c^3) I_1(1,4,1)-8 m_b I_1(2,1,2)+16 m_b q^2 I_1(2,1,3)+(-16 m_b^3+8 m_cm_b^2-8 m_c^2 m_b
\\ \nnb &+&
   16 m_c^3) I_1(2,3,1)+(16 m_b-40 m_c) I_1(3,1,1)+(-24 m_b^3+16 q^2 m_b) I_1(3,1,2)+(-8 m_b^5+16 q^2 m_b^3
 \\ \nnb &-&
   8 q^4 m_b) I_1(3,1,3)+(-16 m_b^3+8 m_c m_b^2-8 m_c^2 m_b+16m_c^3) I_1(3,2,1)+(-8 m_b^5+8 m_c m_b^4+16 m_c^2 m_b^3
   \\ \nnb &-&
   16 m_c^3 m_b^2-8 m_c^4m_b+8 m_c^5) I_1(3,3,1)+(72 m_b^2 m_c-72 m_b^3) I_1(4,1,1)+(-16 m_b
   \\ \nnb &+&
   8 m_c) I_2(1,1,3)-8 m_cI_2(1,2,2)+(16 m_c^3+16 q^2 m_c )I_2(1,2,3)+(-16 m_b-72 m_c) I_2(1,3,1)+(16 m_c^3
   \\ \nnb &+&
   8 q^2 m_c) I_2(1,3,2)+(8 m_c^5+16 q^2m_c^3) I_2(1,3,3)+(-72 m_c^3+48 m_b m_c^2) I_2(1,4,1)-8 m_bI_2(2,1,2)
\\ \nnb &+&
   32 m_b q^2 I_2(2,1,3)+(-32m_b^3+16 m_c m_b^2-16 m_c^2 m_b+32 m_c^3) I_2(2,3,1)+(40 m_b-128 m_c)I_2(3,1,1)
\\ \nnb &+&
   (-40 m_b^3+24 q^2 m_b)I_2(3,1,2)+(-16 m_b^5+32 q^2 m_b^3-16 q^4 m_b)I_2(3,1,3)+(-32 m_b^3+16 m_c m_b^2
\\ \nnb &-&
   16 m_c^2 m_b+32 m_c^3) I_2(3,2,1)+(-16m_b^5+16 m_c m_b^4+32 m_c^2 m_b^3-32 m_c^3 m_b^2
\\  &-&
   16 m_c^4 m_b+16 m_c^5)I_2(3,3,1)+(-144 m_b^3+216 m_c m_b^2) I_2(4,1,1)\Bigg\}\eea

where
 \bea D_i^j \left[I_n(M_1^2,M_2^2)\right]&=& ( M_1^2)^i (M_2^2)^j \frac{\partial_i}{\partial( M_1^2)^i} \frac{\partial^j}
{\partial( M_2^2 )^j}\left[( M_1^2)^i (M_2^2 )^j I_n(M_1^2,M_2^2) \right]~.
\nnb \eea

\begin{thebibliography}{9}



\bibitem{ATLAS}
 G. Aad et al. (ATLAS Collaboration), Phys. Lett. B 716, 1 (2012).
\bibitem{CMS}
 S. Chatrchyan et al. (CMS Collaboration), Phys. Lett. B 716, 30 (2012).
\bibitem{LHCb} R. Aaij et al. (LHCb Collaboration), Phys. Rev. Lett. 108, 251802 (2012); Phys
\bibitem{Wang:2013cha}
  Z.~G.~Wang,
  Eur.\ Phys.\ J.\ C {\bf 73}, 2559 (2013)
  [arXiv:1306.6160 [hep-ph]].

\bibitem{ref2} Wang Zhi-Gang, Commun. Theor. Phys.61,81 (2014).

\bibitem{Buchalla:1998ba}
  G.~Buchalla and A.~J.~Buras,
  Nucl.\ Phys.\ B {\bf 548}, 309 (1999)
  [hep-ph/9901288].
\bibitem{Xiao:2002ew}
  Z.~-j.~Xiao and L.~-p.~Yao,
  Commun.\ Theor.\ Phys.\  {\bf 38}, 683 (2002)
  [hep-ph/0212008].
\bibitem{Buchalla:2010jv}
  G.~Buchalla,
  Nucl.\ Phys.\ Proc.\ Suppl.\  {\bf 209}, 137 (2010)
  [arXiv:1010.2674 [hep-ph]].
%
\bibitem{Geng:2001vy}
  C.~Q.~Geng, C.~W.~Hwang and C.~C.~Liu,
  Phys.\ Rev.\ D {\bf 65}, 094037 (2002)
  [hep-ph/0110376].

%
%
%
%
%
\bibitem{Bashiry:2013waa}
  V.~Bashiry,
  arXiv:1305.6535 [hep-ph].
\bibitem{Ghahramany:2013gya}
  N.~Ghahramany and A.~R.~Houshyar,
  Acta Phys.\ Polon.\ B {\bf 44}, no. 9, 1857 (2013).

\bibitem{Gan:2012tt}
  L.~-F.~Gan, Y.~-L.~Liu, W.~-B.~Chen and M.~-Q.~Huang,
  Commun.\ Theor.\ Phys.\  {\bf 58}, 872 (2012)
  [arXiv:1212.4671 [hep-ph]].

\bibitem{Sarac:2013rpa}
  Y.~Sarac, K.~Azizi and H.~Sundu,
  Nucl.\ Phys.\ Proc.\ Suppl.\  {\bf 245}, 164 (2013).

\bibitem{Khodjamirian:2009ys}
  A.~Khodjamirian, C.~.Klein, T.~.Mannel and N.~Offen,
  Phys.\ Rev.\ D {\bf 80}, 114005 (2009)
  [arXiv:0907.2842 [hep-ph]].

\bibitem{Aliev:2006vs}
  T.~M.~Aliev and M.~Savci,
  Eur.\ Phys.\ J.\ C {\bf 47}, 413 (2006)
  [hep-ph/0601267].

\bibitem{Azizi:2008vv}
  K.~Azizi, F.~Falahati, V.~Bashiry and S.~M.~Zebarjad,
  Phys.\ Rev.\ D {\bf 77}, 114024 (2008)
  [arXiv:0806.0583 [hep-ph]].

  \bibitem{Marques de Carvalho:1999ia}
  R.~S.~Marques de Carvalho, F.~S.~Navarra, M.~Nielsen, E.~Ferreira and H.~G.~Dosch,
  Phys.\ Rev.\ D {\bf 60}, 034009 (1999)
  [hep-ph/9903326].
\bibitem{R13} G. Buchalla and A. Buras,
{\it Nucl. Phys.} {\bf B400} (1993) 225;
{\it Phys. Rev.} {\bf D54} (1996) 6782.

\bibitem{R14} T. Inami and C. S. Lim,
{\it Prog. Theor. Phys.} {\bf 65} (1981) 287.
\bibitem{Aliev:2001in}
  T.~M.~Aliev, A.~Ozpineci and M.~Savci,
  Phys.\ Lett.\ B {\bf 506}, 77 (2001)
  [hep-ph/0101066].

%
%
%
%
%
%
%
\bibitem{R7323} V. V. Kiselev, A. K. Likhoded,  A. I. Onishchenko,
 Nucl. Phys.  {\bf B 569}  (2000) 473.

 \bibitem{R7321} J. Schwinger, Phys. Rev. {\bf 82}, 664 (1951).
\bibitem{pdg12}J. Beringer et al. (Particle Data Group), Phys. Rev. D86, 010001 (2012)
%
%
%
\bibitem{Wang:2012kw}
  Z.~-G.~Wang,
  Eur.\ Phys.\ J.\ A {\bf 49} (2013) 131
  [arXiv:1203.6252 [hep-ph]].
%
%

\bibitem{Shifman1} M. A. Shifman, A. I. Vainshtein,  V. I. Zakharov, Nucl.
Phys. {\bf B147}  (1979) 385.

\bibitem{AbdElHady:1998kc}
  A.~Abd El-Hady, M.~A.~K.~Lodhi and J.~P.~Vary,
  Phys.\ Rev.\ D {\bf 59}, 094001 (1999)
  [hep-ph/9807225].
\bibitem{Kiselev:2000jc}
  V.~V.~Kiselev, A.~E.~Kovalsky and A.~I.~Onishchenko,
  Phys.\ Rev.\ D {\bf 64}, 054009 (2001)
  [hep-ph/0005020].
\end{thebibliography}
\end{document}